\documentclass[12pt]{iopart}
\pdfoutput=1
\usepackage{graphicx}
\usepackage{iopams}

\renewcommand{\d}{\rmd}
\newcommand{\dd}[1]{\rmd #1\;}
\newcommand{\ii}{\rmi}

\renewcommand{\mr}[1]{\mathrm{#1}}
\newcommand{\refcite}[1]{\cite{#1}}
\newcommand{\refscite}[1]{\cite{#1}}

\newcommand{\Ret}{\mr{R}}
\newcommand{\Adv}{\mr{A}}
\newcommand{\K}{\mr{K}}
\newcommand{\Ri}{\mr{R}}
\newcommand{\Le}{\mr{L}}

\newcommand{\ee}{\mr{ee}}
\newcommand{\epe}{\mr{epe}}

\newcommand{\Hc}{\mr{H.c.}}

\newcommand{\sgn}{\mr{sgn}}

\begin{document}

\title{Functional renormalization group study of the Anderson--Holstein model}

\author{M A Laakso, D M Kennes, S G Jakobs and V Meden}

\address{Institut f\"ur Theorie der Statistischen Physik, RWTH Aachen, 52056 Aachen, Germany}
\eads{laakso@physik.rwth-aachen.de}

\begin{abstract}
We present a comprehensive study of the spectral and transport properties in the Anderson--Holstein model both in and out of equilibrium using the functional renormalization group (FRG). We show how the previously established machinery of Matsubara and Keldysh FRG can be extended to include the local phonon mode. Based on the analysis of spectral properties in equilibrium we identify different regimes depending on the strength of the electron--phonon interaction and the frequency of the phonon mode. We supplement these considerations with analytical results from the Kondo model. We also calculate the non-linear differential conductance through the Anderson--Holstein quantum dot and find clear signatures of the presence of the phonon mode.
\end{abstract}


\maketitle

\section{Introduction}

The Anderson--Holstein model is widely used to describe electronic transport through individual molecules contacted between two leads: The lowest unoccupied molecular orbital is described as a single, spin-degenerate (in the absence of a magnetic field) level of a quantum dot with a Coulomb repulsion between electrons of opposite spin. In addition, the molecule can vibrate at some characteristic frequency $\omega_0$, and the electrons interact with the polarization field generated by this vibration.

Equilibrium properties, such as linear conductance and spectral density, of the Anderson--Holstein model have been studied perturbatively \cite{mitra04, galperin07}, and nonperturbatively using numerical renormalization group (NRG) techniques \cite{hewson02,cornaglia04}. In the limit of weak coupling to the leads, a Lang--Firsov transformation together with a generalized Schrieffer--Wolff transformation can be used to obtain an effective Kondo model, which allows for a traditional renormalization group treatment \cite{paaske05}.

Using the Meir--Wingreen formula \cite{meir92} it is possible to obtain a simple expression for the non-linear current--voltage characteristics of the model, under the assumption that the spectral function does not change when a bias voltage is applied. This means essentially that the molecule is kept well-grounded by a good contact with one lead whereas the other lead at a finite voltage is very weakly coupled to the molecule. This can be realized in a scanning tunneling microscope (STM) setup where the differential conductance through the STM tip probes the spectral density of the molecule. Due to the above assumption the many-body calculation can be performed in equilibrium as in \refscite{cornaglia04,paaske05}.

Going beyond the STM-like setup the Anderson--Holstein model in a bias voltage driven non-equilibrium steady-state has been studied using real-time diagrammatics \cite{koenig96}, rate equations \cite{braig03,mitra04,koch05}, and slave-boson techniques with the non-crossing approximation \cite{roura-bas13}. In these cases the double occupation of the dot has been forbidden by considering the limit of an infinitely strong Coulomb repulsion, suppressing all charge fluctuations. Some numerical progress in the case of finite Coulomb repulsion has also been made using the imaginary-time non-equilibrium formalism \cite{han06,han10}.

In addition, a related resonant level model with a local phonon mode has been studied using real-time diagrammatic Monte Carlo \cite{schiro09}, variational Lang--Firsov transformation \cite{koch11}, numerical path integral approaches \cite{muehlbacher08,hutzen12}, NRG \cite{eidelstein13}, and scattering states NRG \cite{jovchev13}. In this model also the time evolution towards the non-equilibrium steady state has been studied \cite{albrecht13}. The lack of spin excludes all Kondo-related effects, however. This model admits a simple description of the interplay between electron tunneling and phonon dynamics in terms of the ratio $\omega_0/\Gamma$, where $\Gamma$ is the tunneling rate of the electrons, both limits $\omega_0/\Gamma\gg1$ and $\omega_0/\Gamma\ll1$ being theoretically well controlled \cite{eidelstein13}.

An experimental realization of an Anderson--Holstein quantum dot can be for example a suspended carbon nanotube, where the principal vibration mode of the tube plays the role of a local phonon mode \cite{leturcq09}. The reported strength of the electron--phonon coupling in this case is relatively high --- the ratio between the coupling constant and the phonon frequency, $\lambda/\omega_0$, reaching values in excess of 5 \cite{leturcq09}. Moreover, due to the length of the nanotubes the ratio $\omega_0/\Gamma$ can be of the order of one, a regime which is theoretically challenging. Accessing this calls for a nonperturbative treatment.

In this article we study the spectral and transport properties in the Anderson--Holstein model using the functional renormalization group (FRG) \cite{metzner12}. We show that in equilibrium FRG can be successfully used to approach the regime of intermediate electron--phonon coupling, resulting in a Kondo-like behavior. We also use the non-equilibrium Keldysh formulation of FRG \cite{jakobs10b,metzner12} to study the non-linear electron transport through the Anderson--Holstein quantum dot and compare our results to first order perturbation theory.

This Article is structured in the following way: We introduce the model and express it in terms of the Keldysh Green functions, and briefly describe the FRG in \sref{sec:model}. We present the equilibrium properties, i.e., spectral densities and linear conductance in \sref{sec:eq}, and non-equilibrium spectral densities and differential conductance in \sref{sec:noneq}. We summarize our findings and describe avenues for future research in \sref{sec:conclusions}. Throughout this paper we work with natural units in which $\hbar=k_B=1$.

\section{Model}\label{sec:model}
\subsection{Hamiltonian and effective action}
The system is described by the Anderson--Holstein Hamiltonian
\begin{eqnarray}
  H=\epsilon(n_{\uparrow}+n_{\downarrow})+Un_{\uparrow}n_{\downarrow}+\sum_{r=\Ri,\Le}(H_{r}+H_{r}^{\mr{coup}}) \nonumber \\ \qquad+\omega_{0}b^{\dagger}b+\lambda(b^{\dagger}+b)(n_{\uparrow}+n_{\downarrow}), \\
  H_{r}=\sum_{k_{r},\sigma}\epsilon_{k_{r}}c_{k_{r}\sigma}^{\dagger}c_{k_{r}\sigma}, \\
  H_{r}^{\mr{coup}}=\sum_{k_{r},\sigma}\gamma_{k_{r}\sigma}c_{k_{r}\sigma}^{\dagger}d_{\sigma}+\Hc,
\end{eqnarray}
where
\begin{eqnarray}
  n_{\sigma}=d_{\sigma}^{\dagger}d_{\sigma},\\
  \epsilon=eV_\mr{G}-\frac{U}{2}+U_\epe,\\
  U_\epe=\frac{2\lambda^2}{\omega_0}.
\end{eqnarray}
The fermionic operators $d_{\sigma}$ and $c_{k_{r}\sigma}$ refer to electrons on the dot and in reservoir $r=\Ri,\Le$, respectively, while the bosonic operator $b$ refers to phonons with frequency $\omega_{0}$ and coupling constant $\lambda$. Furthermore, $eV_\mr{G}$ is the gate voltage and $U$ the strength of the repulsive interaction between electrons on the dot. With our definition of $\epsilon$, $eV_\mr{G}=0$ corresponds to half filling of the dot.

Since we consider only equilibrium, as well as steady state non-equilibrium, the system becomes translationally invariant in time and it is sufficient to work in frequency space. We treat the reservoirs in the wide band limit with a constant density of states $D_{r}$ and assume that the coupling between the reservoir and the dot states does not depend on momentum and spin,
\begin{eqnarray}
  \sum_{k_{r}}\delta(\omega-\epsilon_{k_{r}})=D_{r},\\
  \gamma_{k_{r}\sigma}=\gamma_{r}.
\end{eqnarray}
After integrating out the reservoirs, held locally in equilibrium with chemical potential $\mu_r$ and temperature $T_r$, the noninteracting but reservoir dressed electron propagator on the dot reads
\begin{eqnarray}
  G_{0}(\omega,&\omega')=2\pi\delta(\omega-\omega')G_{0}(\omega),\\
  G_{0}^{\Ret}(\omega)&=\frac{1}{\omega-\epsilon-\sum_{r}\Sigma_{r}^{\Ret}},\nonumber\\
  &=\frac{1}{\omega-\epsilon+\ii\Gamma},\\
  G_{0}^{\K}(\omega)&=G_{0}^{\Ret}(\omega)\left(\sum_{r}\Sigma_{r}^{\K}(\omega)\right)G_{0}^{\Adv}(\omega),
\end{eqnarray}
with\footnote{We note that sometimes $\Gamma$ is defined in the literature with an additional prefactor of $2$.}
\begin{equation}
  \Gamma=\sum_{r}\Gamma_{r},\qquad\Gamma_{r}=\pi D_{r}\gamma_{r}^{\ast}\gamma_{r},
\end{equation}
and
\begin{eqnarray}
  \Sigma_{r}^{\Ret}=-\ii\Gamma_{r},\\
  \Sigma_{r}^{\K}(\omega)=-2\ii\left[1-2n_{r}(\omega)\right]\Gamma_{r},\\
  n_{r}(\omega)=\frac{1}{\rme^{(\omega-\mu_{r})/T_{r}}+1}.
\end{eqnarray}
We present the derivation in terms of the non-equilibrium Keldysh Green functions. Analogous but somewhat simpler expressions hold for the equilibrium Matsubara Green functions \cite{karrasch08}.

Finally, we integrate out the phonons to obtain an effective action for the interacting electrons on the dot:
\begin{equation}
  \fl S[\bar{\psi},\psi]=\bar{\psi}_{1'}(G_0^{-1})_{1'1}\psi_{1}-\frac{1}{4}\bar{\nu}_{1'2'12}^{\ee}\bar{\psi}_{1'}\bar{\psi}_{2'}\psi_2\psi_1-\frac{1}{4}\bar{\nu}_{1'2'12}^{\epe}\bar{\psi}_{1'}\bar{\psi}_{2'}\psi_2\psi_1,
\end{equation}
where the multi-indices $1=(l_1,\omega_1,\sigma_1)$ contain the Keldysh index\footnote{We use the convention where the $2\times2$ matrix Green function is given by $G=\left(\begin{array}{cc} 0 & G^\Adv \\ G^\Ret & G^\K \end{array}\right)$ and the self-energy by $\Sigma=\left(\begin{array}{cc} \Sigma^\K & \Sigma^\Ret \\ \Sigma^\Adv & 0 \end{array}\right)$.}, frequency, and spin, and repeated indices are summed or integrated over: $\sum_l\sum_\sigma\int\frac{\dd{\omega}}{2\pi}$. The direct antisymmetrized electron--electron interaction vertex is given by
\begin{eqnarray}
  \fl \bar{\nu}_{1'2'12}^{\ee}=2\pi\delta(\omega_{1'}+\omega_{2'}-\omega_{1}-\omega_{2})\frac{U}{2}\left(\delta_{\sigma_{1'}\sigma_{1}}\delta_{\sigma_{2'}\sigma_{2}}-\delta_{\sigma_{1'}\sigma_{2}}\delta_{\sigma_{2'}\sigma_{1}}\right)\delta_{\bar{\sigma}_{1}\sigma_{2}} \nonumber \\
  \times\left(\begin{array}{cc}\left(\begin{array}{cc}0 & 1\\
    1 & 0
    \end{array}\right)_{l{}_{1'}l_{1}} & \left(\begin{array}{cc}1 & 0\\
    0 & 1
    \end{array}\right)_{l{}_{1'}l_{1}}\\
    \left(\begin{array}{cc}1 & 0\\
    0 & 1
    \end{array}\right)_{l{}_{1'}l_{1}} & \left(\begin{array}{cc}0 & 1\\
    1 & 0
    \end{array}\right)_{l{}_{1'}l_{1}}
  \end{array}\right)_{l_{2'}l_{2}},\label{eq:nu_ee}
\end{eqnarray}
and the retarded phonon-mediated electron--electron interaction vertex (electron--phonon--electron vertex) by
\begin{eqnarray}
  \fl\bar{\nu}_{1'2'12}^{\epe}=2\pi\delta(\omega_{1'}+\omega_{2'}-\omega_{1}-\omega_{2})\frac{\lambda^{2}}{2} \nonumber \\
  \times\left[\delta_{\sigma{}_{1'}\sigma_{1}}\delta_{\sigma_{2'}\sigma_{2}}\left(\begin{array}{cc}\left(\begin{array}{cc}0 & D\\
    D & 0
    \end{array}\right)_{l_{1'}l_{1}} & \left(\begin{array}{cc}D^{\ast} & K\\
    K & D^{\ast}
    \end{array}\right)_{l_{1'}l_{1}}\\
    \left(\begin{array}{cc}D^{\ast} & K\\
    K & D^{\ast}
    \end{array}\right)_{l_{1'}l_{1}} & \left(\begin{array}{cc}0 & D\\
    D & 0
    \end{array}\right)_{l_{1'}l_{1}}
    \end{array}\right)_{l_{2'}l_{2}}(\Delta)\right.\nonumber \\
    \left.-\delta_{\sigma_{1'}\sigma_{2}}\delta_{\sigma_{2'}\sigma_{1}}\left(\begin{array}{cc}\left(\begin{array}{cc}0 & D\\
    D^{\ast} & K
    \end{array}\right)_{l_{1'}l_{1}} & \left(\begin{array}{cc}D^{\ast} & K\\
    0 & D
    \end{array}\right)_{l_{1'}l_{1}}\\
    \left(\begin{array}{cc}D & 0\\
    K & D^{\ast}
    \end{array}\right)_{l_{1'}l_{1}} & \left(\begin{array}{cc}K & D^{\ast}\\
    D & 0
    \end{array}\right)_{l_{1'}l_{1}}
  \end{array}\right)_{l_{2'}l_{2}}(X)\right],\label{eq:phonon_induced_interaction}
\end{eqnarray}
with the phonon propagators
\begin{eqnarray}
  D(\omega)=\frac{2\omega_{0}}{(\omega+\ii\eta)^{2}-\omega_{0}^{2}},\\
  K(\omega)=-2\pi \ii\left[\delta(\omega-\omega_{0})+\delta(\omega+\omega_{0})\right]\left(1+2n_\mr{p}\right),\label{eq:delta}\\
  n_\mr{p}=\frac{1}{\rme^{\omega_{0}/T_\mr{p}}-1}.
\end{eqnarray}
Here, $T_\mr{p}$ is the phonon temperature and $\eta$ a positive infinitesimal, which also defines the width of the delta peaks in \eref{eq:delta}. In the numerical calculations we assign it a finite value which is small enough not to affect the results. We have also defined the bosonic frequencies
\begin{eqnarray}
 \Pi=\omega_{1}+\omega_{2}=\omega_{1'}+\omega_{2'}, \\
 X=\omega_{2'}-\omega_{1}=\omega_{2}-\omega_{1'}, \\
 \Delta=\omega_{1'}-\omega_{1}=\omega_{2}-\omega_{2'},
\end{eqnarray}
where the second equalities hold due to frequency conservation.

\subsection{Functional renormalization group}

After integrating out exactly both the leads and the phonons we tackle the remaining problem of the interacting electrons on the dot with the functional renormalization group \cite{metzner12}. For the non-equilibrium calculations we use the Keldysh formulation developed for the single-impurity Anderson model (SIAM) in \refcite{jakobs10b}. We use the hybridization to a structureless auxiliary lead, $\Lambda$, as the flow parameter and truncate the hierarchy of FRG equations at the second order. This includes all interactions at least to fourth order in $\lambda$ as well as contributions beyond plain perturbation theory by virtue of the RG flow. We also use the approximate mixing of the channels and the static self-energy feedback described in sections VI B and VI D of \refcite{jakobs10b}, respectively.

Since the Keldysh structure of the bare electron--phonon--electron vertex is compatible with the structure used in the flow equations of \refcite{jakobs10b}, only the initial conditions of the flow need to be adjusted to account for the phonon-mediated interaction. Compared to the flow without phonons, we need to add $\frac{\lambda^{2}}{2}D(\Delta)$ to the initial value of $(a_{\Lambda}^{\d})_{\sigma\bar{\sigma}}(\Delta)$ and of $(a_{\Lambda}^{\d})_{\sigma\sigma}(\Delta)$, and to add $\frac{\lambda^{2}}{2}K(\Delta)$ to the initial value of $(b_{\Lambda}^{\d})_{\sigma\bar{\sigma}}(\Delta)$ and of $(b_{\Lambda}^{\d})_{\sigma\sigma}(\Delta)$; for the definitions of the functions $(a_{\Lambda}^{\d})_{\sigma\sigma'}(\Delta)$ and $(b_{\Lambda}^{\d})_{\sigma\sigma'}(\Delta)$, see Appendix A of \refcite{jakobs10b}. The initial values of $U_{\Lambda}^{\d}$ and $W_{\Lambda\sigma}^{\d}$ change according to (87b) and (87c) of \refcite{jakobs10b}. For the self-energy we obtain the initial conditions
\begin{eqnarray}
\Sigma_{\Lambda=\infty}^{\Ret} & =\left(\frac{U}{2}-U_\epe\right)\delta_{\sigma'_{1}\sigma_{1}},\\
\Sigma_{\Lambda=\infty}^{\K} & =0,
\end{eqnarray}
where the difference to (39c) in \refcite{jakobs10b} results from the Hartree--Fock diagram with the interaction vertex $\bar{\nu}^\epe$. The resulting system of differential equations is integrated from $\Lambda=\infty$ to $\Lambda=0$ numerically.

For the calculations of equilibrium quantities, namely effective mass (\sref{sec:effmass}) and linear conductance (\sref{sec:conductance}), we also use the Matsubara formulation of FRG. To include the local phonon mode we make analogous changes as described above to the initial conditions presented in \refcite{karrasch08}. In addition, with Matsubara FRG we use frequency cut-off as the flow parameter to simplify the calculations.

Once both the self-energy and the vertex function are determined at the end of the FRG flow, we obtain the full interacting spectral function via
\begin{equation}
  \rho_\sigma(\omega)=\frac{1}{\pi}\frac{\Gamma-\mr{Im}\Sigma^\Ret_\sigma(\omega)}{[\omega-\epsilon-\mr{Re}\Sigma^\Ret_\sigma(\omega)]^2+[\Gamma-\mr{Im}\Sigma^\Ret_\sigma(\omega)]^2},
\end{equation}
and the current by \cite{meir92}
\begin{equation}\label{eq:current}
  I=2e\int\dd{\omega}\left[n_\Le(\omega)-n_\Ri(\omega)\right]\frac{\Gamma_\Le\Gamma_\Ri}{\Gamma}\sum_\sigma\rho_\sigma(\omega).
\end{equation}
The (differential) conductance is then obtained by numerically differentiating the current with respect to the bias voltage.

\subsection{First order perturbation theory}\label{sec:1pt}

For $U=0$ and small $U_\epe/\Gamma$ we compare FRG results with a perturbation theory to first order in $U_\epe$. The two addends (Hartree and Fock) of $\bar{\nu}^\epe$ in \eref{eq:phonon_induced_interaction} yield two contributions to the retarded self-energy in first order perturbation theory. The first addend is frequency independent,
\begin{equation}
  \Sigma_{\mr{H}\,\sigma'\sigma}^\Ret=-\delta_{\sigma'\sigma}U_\epe(\bar{n}_{\uparrow}+\bar{n}_{\downarrow}),
\end{equation}
where $\bar{n}_{\sigma}$ denotes the average occupation of the spin state $\sigma$ in the noninteracting dot. We consider it reasonable to evaluate this addend self-consistently in order to maintain particle-hole symmetry. Then it satisfies
\begin{equation}
  \Sigma_{\mr{H}}^{\Ret}=-U_{\epe}\frac{2}{\pi}\int \dd{\omega}n_{\mr{eff}}(\omega)\frac{\Gamma}{\left(\omega-\epsilon-\Sigma_{\mr{H}}^{\Ret}\right)^2+\Gamma^2},
\end{equation}
where we have suppressed the spin index and used
\begin{equation}
  n_{\mr{eff}}(\omega)=\frac{1}{\Gamma}\left[\Gamma_\Le n_\Le(\omega)+\Gamma_\Ri n_\Ri(\omega)\right].
\end{equation}
In the special case $T_\Le=T_\Ri=0$ follows
\begin{equation}
  \Sigma_{\mr{H}}^{\Ret}=U_{\epe}\left(-1+\frac{2}{\pi\Gamma}\sum_{r}\Gamma_{r}\arctan\frac{\epsilon+\Sigma_{\mr{H}}^{\Ret}-\mu_{r}}{\Gamma}\right).
\end{equation}
Consider this equation for $\mu_\Le=\mu_\Ri=\mu$. Then it has a unique solution for $U_{\epe}/\Gamma\le\pi/2$. In particular at $eV_\mr{G}=\mu$ the solution is $\Sigma_{\mr{H}}^{\Ret}(eV_\mr{G}=\mu)=-U_{\epe}$. For $U_{\epe}/\Gamma>\pi/2$ and $V_\mr{G}$ sufficiently close to $\mu$ the equation has three solutions and leads to a hysteresis of $\Sigma_{\mr{H}}^{\Ret}(V_\mr{G})$. Therefore we use the self-consistent approach only for $U_{\epe}/\Gamma\le\pi/2$.

The second addend in \eref{eq:phonon_induced_interaction} yields in first order perturbation theory the contribution
\begin{equation}
  \fl \Sigma_{\mr{F}\,\sigma'\sigma}^{\Ret}(\omega)=\frac{\ii}{4\pi}\lambda^2\delta_{\sigma'\sigma}\int \dd{\omega'}\left[K(\omega'-\omega)G_{0\,\sigma}^{\Ret}(\omega')+D^{\ast}(\omega'-\omega)G_{0\,\sigma}^{\K}(\omega')\right].
\end{equation}
Here we choose the free propagation to include the self-consistent Hartree contribution determined above. For $T_\Le=T_\Ri=0$ the integral can be evaluated analytically, yielding
\begin{eqnarray}
  \fl \Sigma_{\mr{F}}^{\Ret}(\omega)=\frac{\lambda^2}{2}\left\{\sum_{s=\pm1}\frac{1}{\omega+s\omega_{0}-\tilde{\epsilon}+\ii\Gamma}\right. \nonumber \\
-\frac{1}{\Gamma}\sum_{r}\sum_{s,s'=\pm1}\frac{s\Gamma_r}{\omega+s\omega_{0}-\tilde{\epsilon}+\ii s'\Gamma}\left[\frac{1}{\pi}\arctan\frac{\tilde{\epsilon}-\mu_{r}}{\Gamma}\right.\nonumber \\ \left.\left.+\frac{s'}{2}\sgn(\omega+s\omega_{0}-\mu_{r})+\frac{\ii s'}{\pi}\ln\left|\frac{\omega+s\omega_{0}-\mu_{r}+\ii\eta}{\tilde{\epsilon}-\mu_{r}-\ii s'\Gamma}\right|\right]\right\},
\end{eqnarray}
where we have again suppressed the spin index and set 
\begin{equation}
  \tilde{\epsilon}=\epsilon+\Sigma_{\mr{H}}^{\Ret}.
\end{equation}
The total retarded self-energy in first order perturbation theory is
\begin{equation}
  \Sigma_{\mr{1PT}}^{\Ret}(\omega)=\Sigma_{\mr{H}}^{\Ret}+\Sigma_{\mr{F}}^{\Ret}(\omega).
\end{equation}
We note that $\Sigma_{\mr{F}}^{\Ret}(\omega)$ diverges logarithmically at $\omega=\pm\omega_{0}+\mu_{r}=\pm\omega_{0}\pm eV/2$. Consequently, the spectral density in first order perturbation theory vanishes at these frequencies. An exception is the particle-hole symmetric situation where the $eV=2\omega_0$ dip at zero energy is missing (cf.~\fref{fig:spectralbias} and \fref{fig:spectralbias3}). This is due to a cancellation of these logarithmic divergencies.

\subsection{Choice of the parameters}
The Anderson--Holstein model has a large number of physical parameters which can be varied, thus making a thorough study of the model complicated. To keep things manageable, we restrict our study in this Article to the case of a vanishing magnetic field, and zero temperature, $T_\Le=T_\Ri=T_\mr{p}=0$. Moreover, to enhance the visibility of phonon-related effects, we will focus on the case where the electron--phonon mediated interaction dominates over the bare Coulomb interaction and therefore mainly set $U=0$. We will also restrict our study to a left-right symmetric structure, $\Gamma_\Le=\Gamma_\Ri$. Using the tunneling rate to the leads, $\Gamma$, as a characteristic energy scale we are left with the freedom to vary the strength of the electron--phonon coupling, $\lambda/\Gamma$, and the frequency of the phonon mode, $\omega_0/\Gamma$. Furthermore, it turns out (see \sref{sec:renormalization}) that the physically most relevant combinations are $\lambda/\omega_0$ and $2\lambda^2/(\omega_0\Gamma)\equiv U_\epe/\Gamma$. We will also study how the results change when the average occupation of the dot is changed by tuning the gate voltage $eV_\mr{G}$.

\section{Equilibrium properties}\label{sec:eq}

\subsection{Identification of the parameter regimes}\label{sec:renormalization}

The spinless resonant level model with a local phonon mode is commonly characterized in terms of the ratio $\omega_0/\Gamma$ \cite{eidelstein13}: In the \textit{adiabatic} regime, $\omega_0/\Gamma\ll1$, the local phonon is too slow to effectively respond to the rapidly tunneling electrons and its presence has little effect on their motion. In the \textit{anti-adiabatic} regime, $\omega_0/\Gamma\gg1$, an electron on the dot can lower its energy by forming a polaron with the phonon mode, the energy shift being $E_\mr{p}=-2\lambda^2/\omega_0=-U_\epe$. In the limit $\lambda/\omega_0\gg1$ the tunneling rate becomes strongly renormalized to $\Gamma_\mr{eff}=\Gamma \rme^{-(\lambda/\omega_0)^2}\ll\Gamma$. For this reason it is expected that for strong electron--phonon coupling the crossover from anti-adiabatic to the adiabatic regime happens at $\omega_0/\Gamma_\mr{eff}\approx1$ instead of $\omega_0/\Gamma\approx1$. For the resonant level model this has been confirmed using NRG in \refcite{eidelstein13}.

In the Anderson--Holstein model the presence of the spin leads to richer physics than what can be simply described by an effective tunneling rate. However, the width of the central Kondo resonance in the spectral function is similarly an important quantity. The renormalization of the width can be described by $\Gamma_\mr{ren}=\Gamma/m^\ast$ with the effective mass, $m^\ast=1-\left.\partial_\omega\Sigma^\Ret_\sigma\right|_{\omega=0}$. To leading order in $\Gamma$, analytical estimates can be derived from an effective Kondo model \cite{cornaglia04}.

In the limit $U_\epe\ll U$, the Kondo coupling constant is given by \cite{cornaglia04}
\begin{equation}
  J_KD\approx\frac{8\Gamma}{\pi U}\left(1+\frac{\frac{U_\epe}{U}}{1+\frac{U}{2\omega_0}}\right),
\end{equation}
and the effective mass by
\begin{equation}
  m^\ast\propto\exp\left(\frac{1}{J_KD}\right)=\exp\left[\frac{\pi U}{8\Gamma}\left(1-\frac{\frac{U_\epe}{U}}{1+\frac{U}{2\omega_0}}\right)\right],
\end{equation}
i.e., the electron--phonon coupling reduces the sharpening of the Kondo resonance. This is especially intuitive in the limit $\omega_0\gg U$, where the effective electron--electron interaction is given simply by $U_\mr{eff}=U-U_\epe$.

We are more interested in the limit $U_\epe\gg U$. In this case the effective model is the anisotropic Kondo model with the coupling constants
\begin{equation}
  J_{\parallel,\perp}D\approx\frac{4\Gamma}{\pi \omega_0}\rme^{-(\lambda/\omega_0)^2}\gamma_\pm([\lambda/\omega_0]^2,U/[2\omega_0]),
\end{equation}
where
\begin{equation}
  \gamma_\pm(x,y)=\int_0^1\dd{t}t^{x-y-1}\rme^{\pm xt}.
\end{equation}
For $J_\perp\ll J_\parallel$,
\begin{equation}
  m^\ast\propto\left(\frac{J_\parallel}{J_\perp}\right)^\frac{1}{J_\parallel D},
\end{equation}
and for $J_\perp\approx J_\parallel$
\begin{equation}
  m^\ast\propto\exp\left(\frac{1}{J_\parallel D}\right).
\end{equation}
When $\lambda\gtrsim\omega_0,U$ we find
\begin{equation}
  J_\perp D\approx\frac{4\Gamma}{\pi\lambda}\sqrt{2\pi}\rme^{-2(\lambda/\omega_0)^2+1}\left[1-\left(\frac{\omega_0}{\lambda}\right)^2\right]^{(\lambda/\omega_0)^2-1},
\end{equation}
and
\begin{equation}
  J_\parallel D\approx\frac{4\Gamma}{\pi U_\epe}.
\end{equation}
For $\lambda\gg\omega_0$ we have $J_\parallel/J_\perp\propto\exp[2(\lambda/\omega_0)^2]$, and
\begin{equation}\label{eq:gammaefflambdafourth}
  m^\ast\propto\exp\left[\frac{\pi\omega_0}{\Gamma}\left(\frac{\lambda}{\omega_0}\right)^4\right].
\end{equation}
On the other hand, in the limit $\omega_0\gg\lambda\gg U$, $J_\perp D=J_\parallel D=8\Gamma/(\pi U_\epe)$, and
\begin{equation}\label{eq:gammaefflambdasquared}
  m^\ast\propto\exp\left(\frac{\pi U_\epe}{8\Gamma}\right),
\end{equation}
which is equal to the effective mass in the regular Anderson model with $U$ replaced by $U_\epe$.

Let us now consider the case where $U=0$. The effective anisotropic Kondo model is still valid when $U_\epe/\Gamma\gg1$. For the opposite limit $U_\epe/\Gamma\ll1$ and weak electron--phonon coupling, $\lambda/\omega_0\ll1$, the problem is perturbative and no Kondo-like exponential behavior can be expected. For $\lambda/\omega_0\gg1$ we find that the effective mass behaves exponentially, but it is given by \eref{eq:gammaefflambdasquared} instead of \eref{eq:gammaefflambdafourth} as for $\lambda/\omega_0\gg1$, $U_\epe/\Gamma\gg1$ (see \sref{sec:effmass}).

\begin{figure}[ht]
  \centering
  \includegraphics[width=1.00\textwidth]{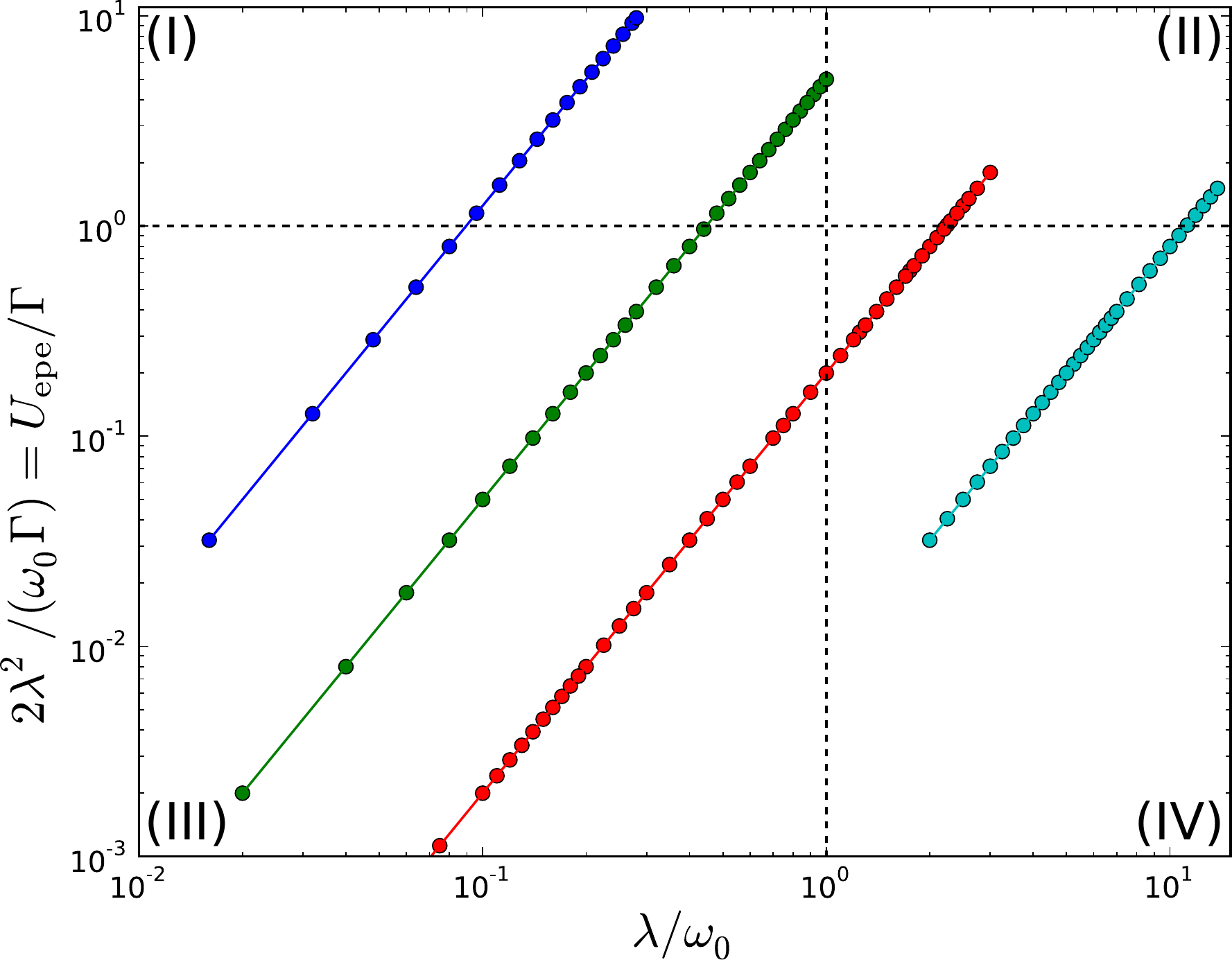}
  \caption{Illustration of the division into four different parameter regimes: attractive single-impurity Anderson model (I), extended anti-adiabatic (II), perturbative (III), and adiabatic (IV). The colored lines and data points correspond to those of \fref{fig:gammaefffrg}.}
  \label{fig:regimes}
\end{figure}
For $\lambda/\omega_0\gg1$, $\omega_0\approx\Gamma$, it follows that $U_\epe/\Gamma\gg1$. This lies in the regime of strong renormalization, regime (II) in \fref{fig:regimes}, and is therefore similar to the extended anti-adiabatic regime of the resonant level model \cite{eidelstein13}. Decreasing $\omega_0/\Gamma$ traces a vertical line towards the adiabatic regime (IV), with the crossover happening roughly at $U_\epe/\Gamma\approx1$ where $\omega_0\ll\Gamma$. Regime (I) we identify with the negative-$U$, i.e., attractive SIAM with $U=-U_\epe$. Finally, in regime (III) the electron--phonon interaction is weak, and the model can be studied perturbatively.

\subsection{Effective mass from FRG}\label{sec:effmass}

\begin{figure}[ht]
  \centering
  \includegraphics[width=1.00\textwidth]{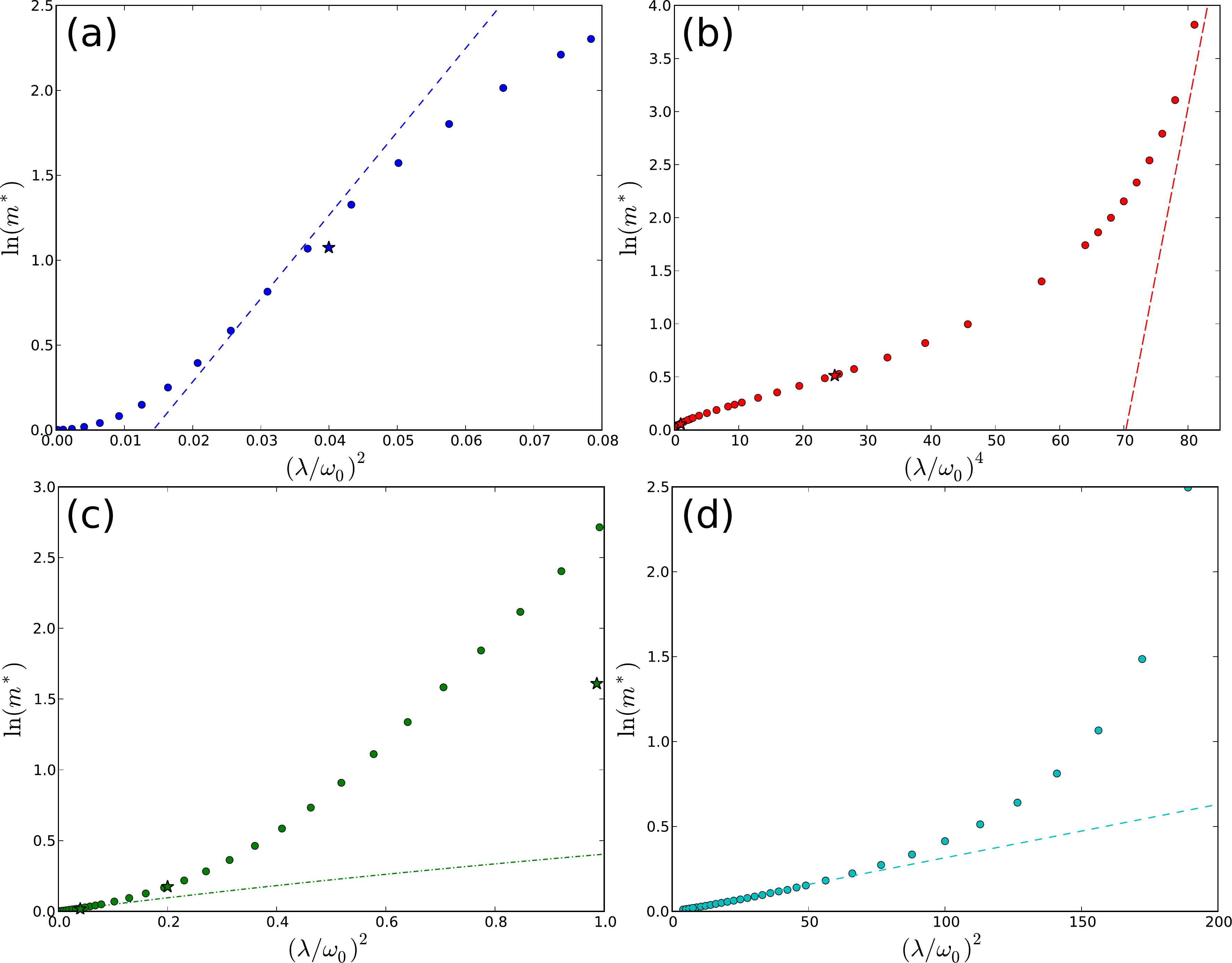}
  \caption{Logarithm of the effective mass as a function of $(\lambda/\omega_0)^2$ or $(\lambda/\omega_0)^4$ for $\omega_0/\Gamma=62.5$ (a), $0.1$ (b), $2.5$ (c), and $0.004$ (d), calculated with the Matsubara FRG. Stars mark data points calculated with the Keldysh FRG. Dashed lines in (a), (b), and (d) are analytical asymptotic predictions from the Kondo model. Dashed-dotted line in (c) corresponds to first order perturbation theory. See text for details.}
  \label{fig:gammaefffrg}
\end{figure}
The dependence of the effective mass on $\lambda/\omega_0$ with $\omega_0/\Gamma$ fixed, obtained numerically from the equilibrium Matsubara FRG, is shown in \fref{fig:gammaefffrg}. A few data points from Keldysh FRG are included for comparison. Varying $\lambda/\omega_0$ in this way traces a line across the four different regimes described above (cf.~\fref{fig:regimes}).

In the regime (I), the effective mass should be described by \eref{eq:gammaefflambdasquared}, indicated as a dashed line in \fref{fig:gammaefffrg}(a). The numerically calculated effective mass for $(\lambda/\omega_0)^2\in [0.02,0.05]$ grows with a slightly smaller prefactor in the exponential function than the expected $\pi\omega_0/(4\Gamma)$, however. A similar discrepancy is seen in the FRG studies of the SIAM \cite{karrasch08,jakobs10b}. For very large $U_\epe/\Gamma$, $(\lambda/\omega_0)^2\gtrsim0.05$, the effective mass bends away from the exponentially increasing trend. This effect is also known from the SIAM, signaling the limit of the validity of our FRG method.

In the regime (IV), the effective mass is well described by \eref{eq:gammaefflambdasquared} as is evident from \fref{fig:gammaefffrg}(d) for $(\lambda/\omega_0)^2\in [4,50]$. As $U_\epe/\Gamma$ approaches unity at $(\lambda/\omega_0)^2=125$ the effective mass starts to increase rapidly. We are unable to reach far into regime (II), with $m^\ast$ described by \eref{eq:gammaefflambdafourth}, since the renormalized interaction vertex grows too large\footnote{We use a criterion where the FRG result is deemed unreliable if the renormalized value of the vertex function grows to more than ten times its original value at any energy during the flow.}, again signaling the limit of the validity of our FRG method. This is even more clear from \fref{fig:gammaefffrg}(b) in which $\ln(m^\ast)$ is plotted against $(\lambda/\omega_0)^4$ instead of $(\lambda/\omega_0)^2$ as in (a), (c), and (d). The behavior described by \eref{eq:gammaefflambdafourth}, indicated as a dashed line, is not observed even for the largest values of $\lambda/\omega_0$. Here, $U_\epe/\Gamma$ reaches unity at $(\lambda/\omega_0)^4=25$. In addition, as we approach regime (II) so that $U_\epe/\Gamma\gg1$, the effective mass from Keldysh FRG starts to deviate from that of Matsubara FRG. This can be clearly seen from the rightmost Keldysh data point in \fref{fig:gammaefffrg}(c).

In the regime (III), c.~f.~\fref{fig:gammaefffrg}(c), the effective mass from FRG and first order perturbation theory agree up to $\lambda/\omega_0\approx0.3$, or, equivalently, $U_\epe/\Gamma\approx0.45$.

\subsection{Linear conductance}\label{sec:conductance}

\begin{figure}[ht]
  \centering
  \includegraphics[width=1.00\textwidth]{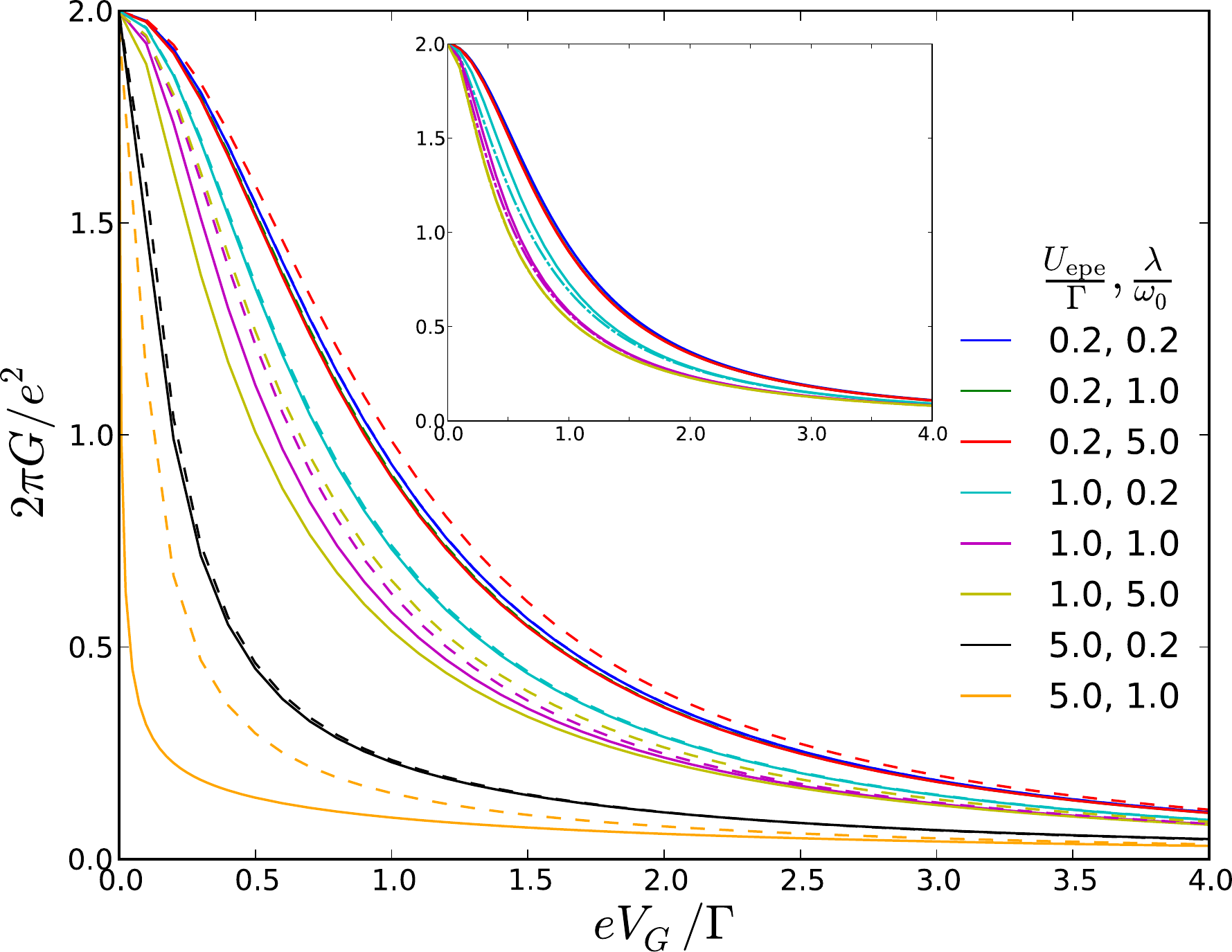}
  \caption{Linear conductance through the Anderson--Holstein quantum dot as a function of the gate voltage for different values of $U_\epe/\Gamma$ and $\lambda/\omega_0$. The lines fall roughly into three groups based on the value of $U_\epe/\Gamma$. Solid lines are obtained with the equiliubrium Matsubara formalism and dashed lines with the non-equilibrium Keldysh formalism. Inset shows a comparison to first order perturbation theory (dash-dotted lines).}
  \label{fig:linearconductance}
\end{figure}
The linear conductance as a function of the gate voltage is shown in \fref{fig:linearconductance}. In general the linear conductance exhibits a peak at a vanishing gate voltage, i.e., at the particle--hole symmetric point. In the case considered here, $U_\epe\gg U$, the low-energy excitations are described by charge fluctuations instead of the spin fluctuations of the ordinary SIAM. As a result the roles of the charge and spin are essentially switched. Application of a gate voltage then has a similar effect as the application of a magnetic field in the ordinary SIAM, leading to a collapse of the conductance plateau. We observe numerically that the width of the resulting conductance peak decreases linearly with increasing $U_\epe/\Gamma$ except when approaching the regime (II) of strong electron--phonon coupling, where the width of the peak decreases significantly faster. In all cases $G=2e^2/h$ for $eV_\mr{G}=0$. 

Looking at the inset of \fref{fig:linearconductance}, results from first order perturbation theory agree perfectly with the results from FRG for $U_\epe/\Gamma=0.2$, whereas for $U_\epe/\Gamma=1.0$ deviations start to appear.

It is important to note that the Keldysh formulation of FRG produces markedly different results than the equilibrium FRG when approaching the regime (II). This is another signature of reaching the validity limit of our truncated FRG equations.

\section{Non-equilibrium properties}\label{sec:noneq}

\subsection{Spectral density}\label{sec:spectral}

To understand the nonlinear current--voltage characteristics of the model, let us first look at how the spectral density changes upon applying a bias voltage.

In general, features in the spectral density appear at various energies. For $eV_\mr{G}=0$, the main resonance lies at $\omega=0$ with phonon steps at $\omega=\pm\omega_0$. The phonon steps are accompanied by shoulders at higher energies with a width that depends on $U_\epe/\Gamma$. At finite bias voltages these steps split to $\omega=\pm\omega_0\pm eV/2$, each with an accompanying shoulder. As shown in \sref{sec:1pt}, the spectral density in first order perturbation theory goes to zero at these energies. The most pronounced effect of the higher order terms included in FRG is the smoothening of these features.

\begin{figure}[ht]
  \centering
  \includegraphics[width=1.00\textwidth]{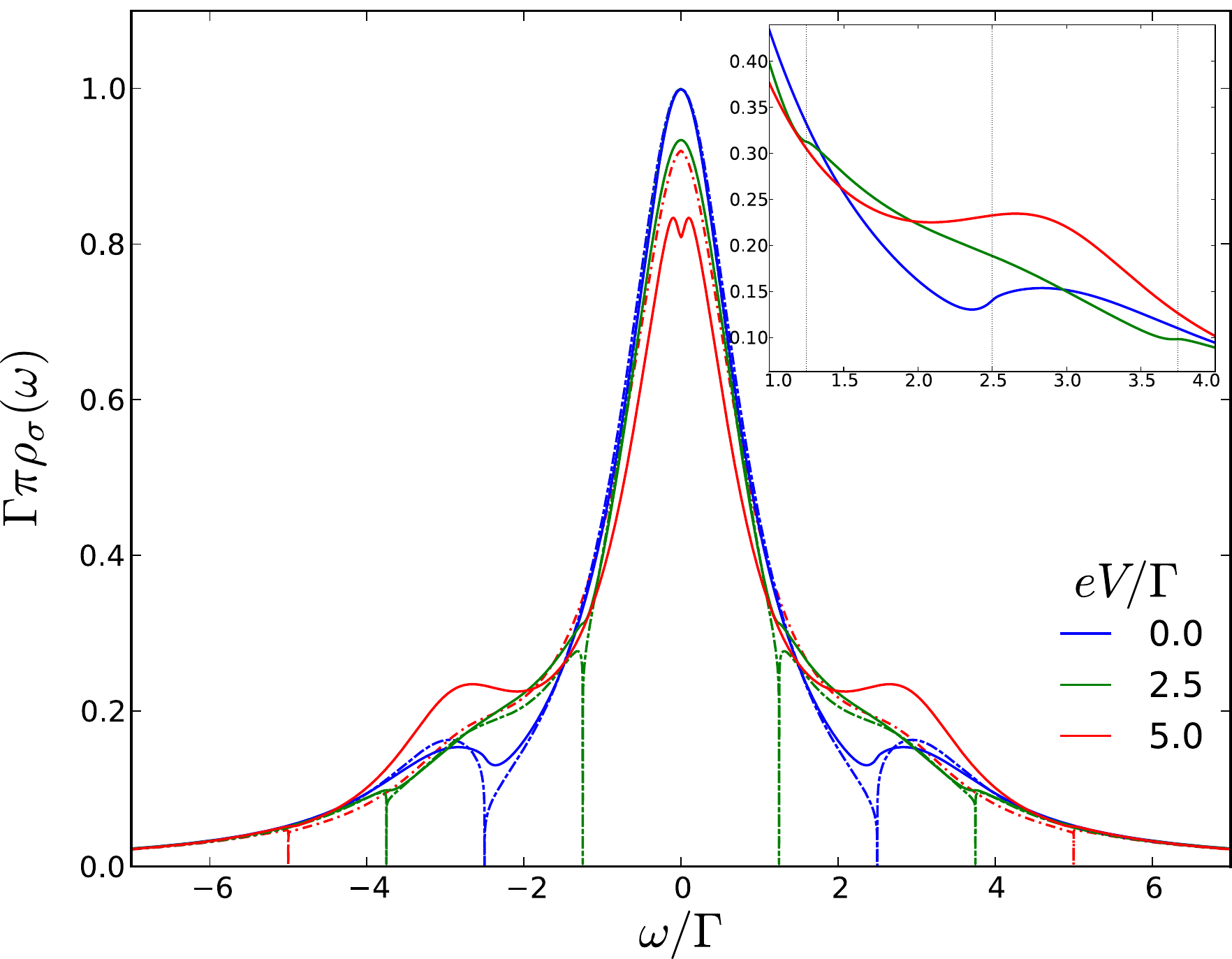}
  \caption{Spectral density of the Anderson--Holstein quantum dot, calculated with the Keldysh FRG for different values of bias voltage and $V_\mr{G}=0$, $\lambda/\omega_0=1/\sqrt{5}\approx0.45$, $U_\epe/\Gamma=1$, resulting in $\omega_0/\Gamma=2.5$. Dash-dotted lines are obtained with first order perturbation theory, peak of the $eV/\Gamma=2.5$ curve being hidden under the zero bias curves. Inset shows a detailed view of the features around $\omega_0$. The grid lines are located at $\omega/\Gamma=1.25$, $2.5$, and $3.75$.}
  \label{fig:spectralbias}
\end{figure}
For $U_\epe/\Gamma\lesssim1$ (see \fref{fig:spectralbias}) the height of the central resonance decreases as the bias voltage is increased, whereas the height of the shoulders increase. When $eV=2\omega_0$, the steps at $\omega=\pm(\omega_0-eV/2)$ merge as can be seen from the double peak structure at zero energy. For $eV>2\omega_0$ these features vanish completely, leaving only the kinks at $\omega=\pm(\omega_0+eV/2)$ (barely visible in the main panel of \fref{fig:spectralbias}).

\begin{figure}[ht]
  \centering
  \includegraphics[width=1.00\textwidth]{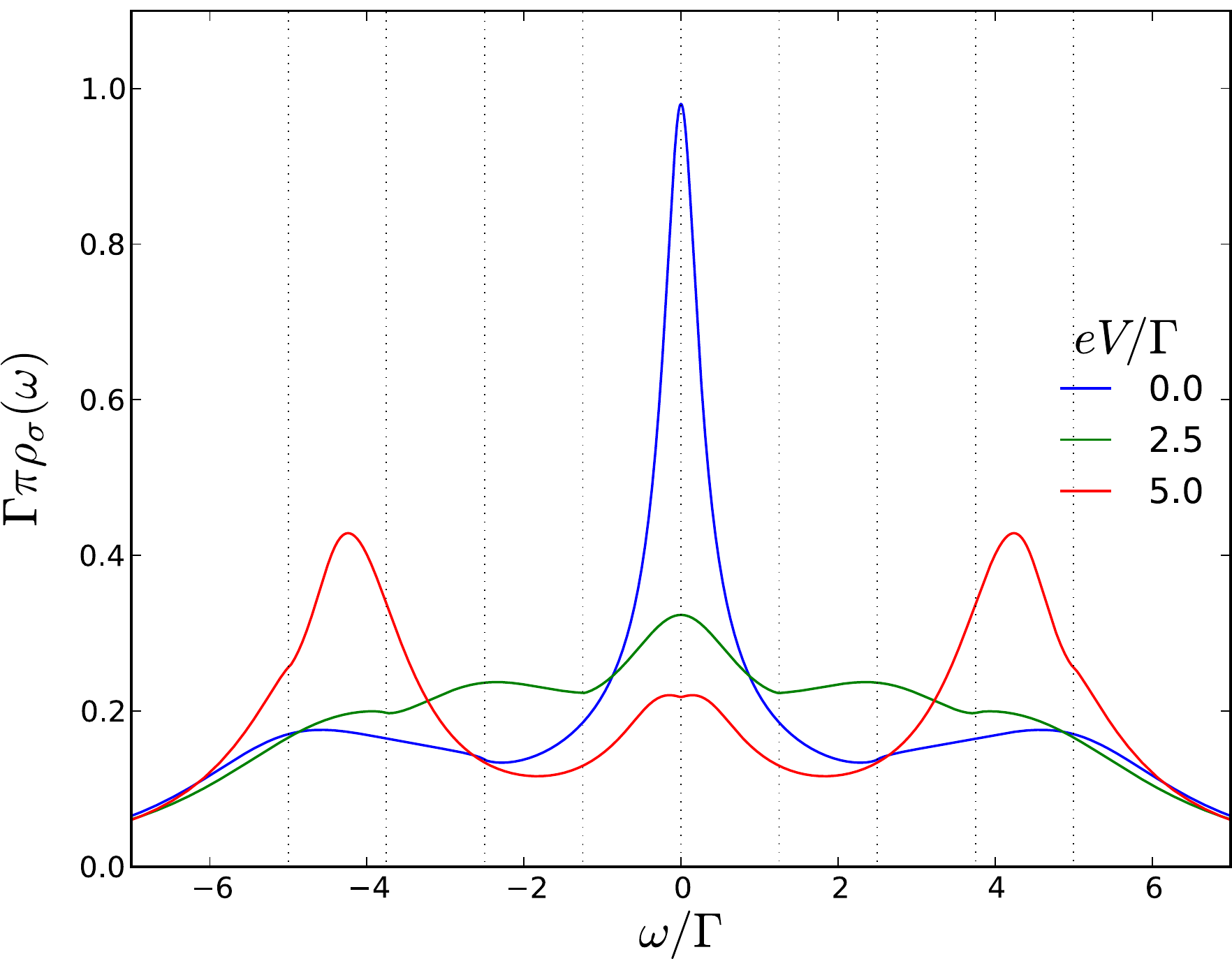}
  \caption{Spectral density of the Anderson--Holstein quantum dot, calculated with the Keldysh FRG for different values of bias voltage and $V_\mr{G}=0$, $\lambda/\omega_0=1$, $U_\epe/\Gamma=5$, resulting in $\omega_0/\Gamma=2.5$. The grid lines are located at $\omega/\Gamma=0.0$, $\pm1.25$, $\pm2.5$, $\pm3.75$, and $\pm5.0$.}
  \label{fig:spectralbias2}
\end{figure}
For $U_\epe/\Gamma\gg1$ (see \fref{fig:spectralbias2}) the spectral weight is significantly redistributed: The height of the central resonance decreases quickly as the bias voltage is increased. The maxima of the shoulders approach $\omega=\pm(\omega_0+eV/2)$ and they increase in height. The behavior of the steps at $\omega=\pm\omega_0\pm eV/2$ is similar as in the case $U_\epe/\Gamma\lesssim1$.

We note that the validity of FRG when approaching regime (II) may be somewhat questionable in view of the discussion in \sref{sec:effmass} and \sref{sec:conductance}. However, since there are no comparable results from any other method available in the literature, we have decided to present non-equilibrium results at the border towards this regime as well. We hope that this motivates further study of the topic by experts using different, suitable strong coupling methods. Moreover, for the ordinary SIAM the differential conductance obtained by FRG agrees with the differential conductance obtained by other numerical methods \cite{eckel10} despite deviations in the spectral function. This fact lends credence to the $U_\epe/\Gamma=5$ results presented in this section.

\begin{figure}[ht]
  \centering
  \includegraphics[width=1.00\textwidth]{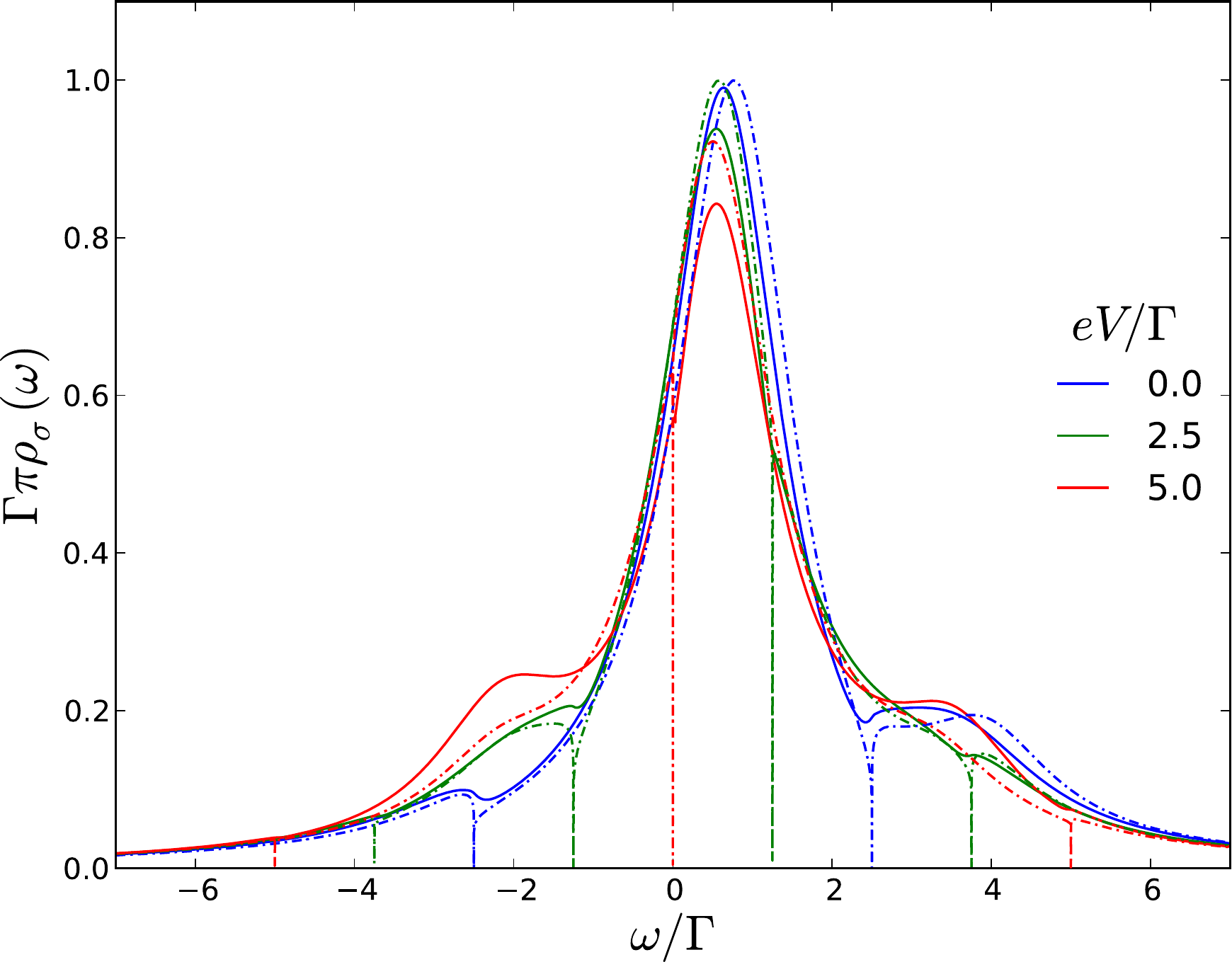}
  \caption{Spectral density of the Anderson--Holstein quantum dot, calculated with the Keldysh FRG for different values of bias voltage and $eV_\mr{G}/\Gamma=0.5$, $\lambda/\omega_0=1/\sqrt{5}\approx0.45$, $U_\epe/\Gamma=1$, resulting in $\omega_0/\Gamma=2.5$. Dash-dotted lines are obtained with first order perturbation theory.}
  \label{fig:spectralbias3}
\end{figure}
Finally, for a finite gate voltage (see \fref{fig:spectralbias3}) the central resonance and the phonon shoulders shift approximately by $eV_\mr{G}$, while the phonon steps stay at the same energies.

\subsection{Differential conductance}
According to \eref{eq:current}, the electric current at $T=0$ and $\mu_{L,R}=\pm eV/2$ is proportional to the integrated spectral weight in the transport window from $\omega=-eV/2$ to $eV/2$. For the differential conductance follows
\begin{equation}\label{eq:diffconductance}
   \fl G_\mr{diff}=\frac{\d I}{\d V}=2e\frac{\Gamma_\Le\Gamma_\Ri}{\Gamma}\sum_{\sigma}
   \left[\frac{e}{2}\sum_{s=\pm1}
   \rho_{\sigma}^{(V)}\left(\omega=s\frac{eV}{2}\right)
   +\int_{-eV/2}^{eV/2}\dd{\omega}\frac{\d}{\d V}
   \rho_{\sigma}^{(V)}(\omega)\right].
\end{equation}
The two addends in square brackets describe two mechanisms how the spectral weight in the transport window changes when $V$ is increased. The first is due to the fact that the transport window gets larger when voltage is increased; it is always positive. The second stems from the fact that the spectral function changes its form when voltage is increased.

\begin{figure}[ht]
  \centering
  \includegraphics[width=1.00\textwidth]{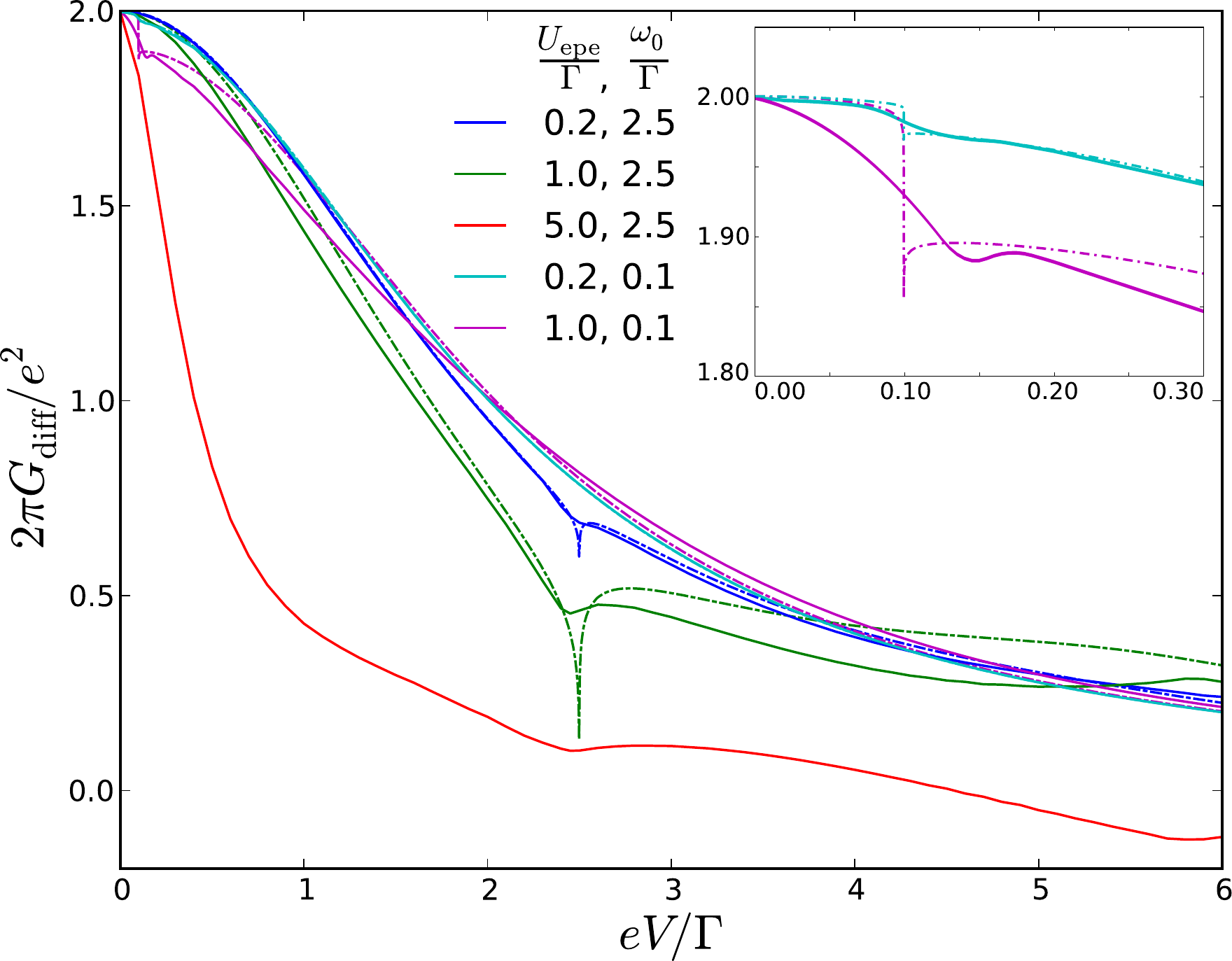}
  \caption{Differential conductance through the Anderson--Holstein quantum dot as a function of bias voltage for $V_\mr{G}=0$, and different values of $U_\epe/\Gamma$ and $\omega_0/\Gamma$, calculated with the Keldysh FRG. Dash-dotted lines are obtained with first order perturbation theory. Inset shows a detailed view of the differential conductance near $eV=\omega_0$ for $\omega_0/\Gamma=0.1$.}
  \label{fig:diffconductance}
\end{figure}
The differential conductance as a function of bias voltage is shown in \fref{fig:diffconductance} for $V_\mr{G}=0$. At a vanishing bias voltage only the first addend contributes, and $\rho_{\sigma}(0)=1/(\pi\Gamma)$ yields $G(V=0)=2e^{2}/h$ (with $h=2\pi$ in our units) as already noted for the linear conductance in \sref{sec:conductance}. When voltage is increased, both contributions lead to a decrease in differential conductance: the first addend decreases, since the value of the spectral function decreases with increasing $|\omega|$ from its maximal value at $\omega=0$; the second addend becomes negative, since spectral weight is moved to higher energies outside the transport window with increasing voltage. For small $U_\epe$ this second effect is not very pronounced (compare \fref{fig:spectralbias}). The width of the peak in the differential conductance at $V=0$ is then essentially determined by the first addend and hence given by twice the width of the central peak in the spectral function. For larger $U_\epe$, the second effect is of considerable importance (compare \fref{fig:spectralbias2}). The width of the peak in the differential conductance at $V=0$ is then significantly smaller than twice the width of the central peak in the spectral function. For large $U_\epe$ and large bias voltage, the negative second contribution may even dominate over the first one. A negative differential conductance results, as seen in \fref{fig:diffconductance} for $U_\epe/\Gamma=5,\omega_0/\Gamma=2.5$ and $eV/\Gamma$ around $5$. At $eV=\omega_0$, the phonon steps of the spectral function situated at $\omega=\pm(\omega_0-eV/2)$ enter the transport window. This leads to the kinks of the differential conductance at $eV=\omega_0$ visible in \fref{fig:diffconductance}.

\begin{figure}[ht]
  \centering
  \includegraphics[width=1.00\textwidth]{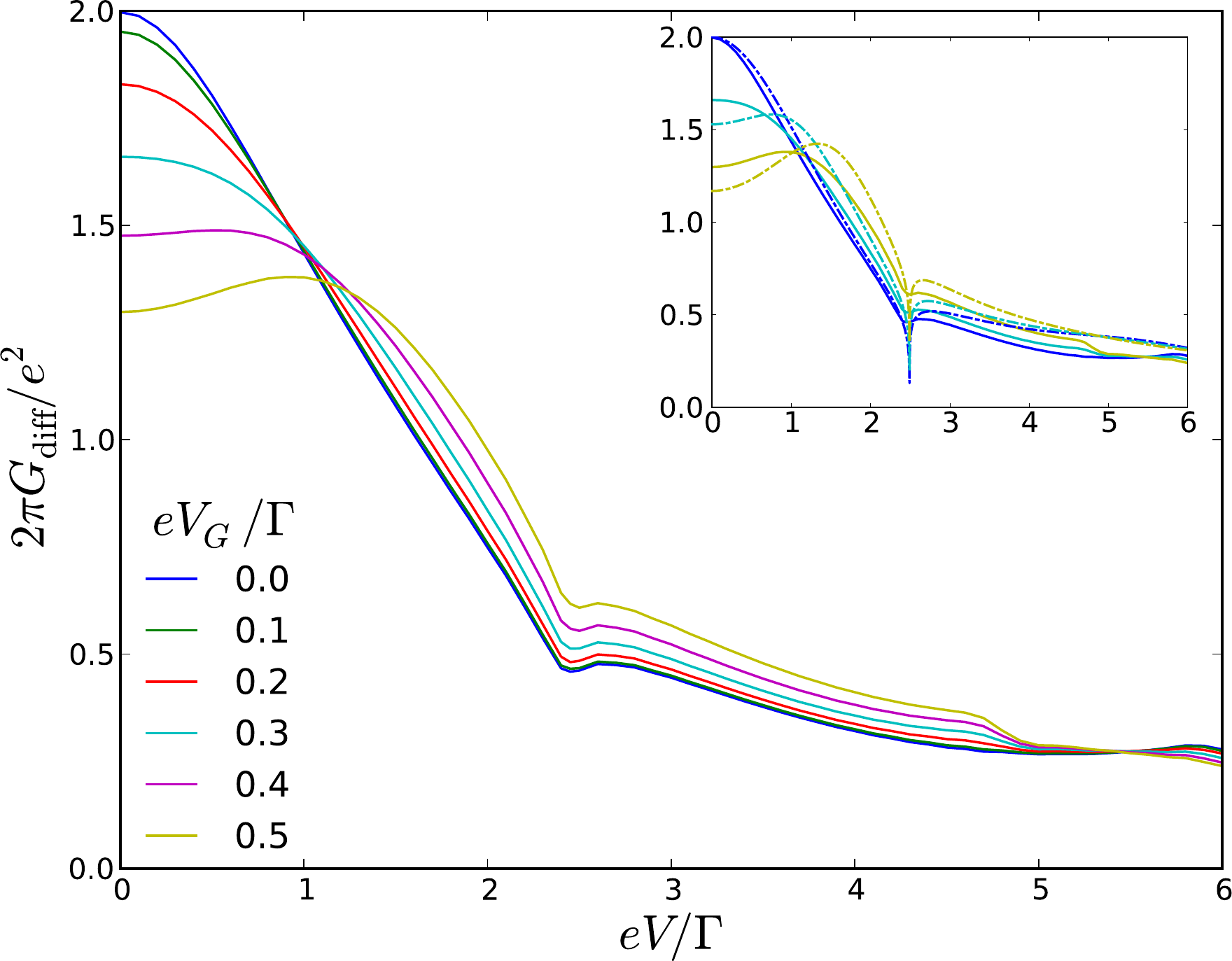}
  \caption{Differential conductance through the Anderson--Holstein quantum dot as a function of bias voltage for different values of gate voltage and $U_\epe/\Gamma=1$, $\omega_0/\Gamma=2.5$, calculated with the Keldysh FRG. Inset shows a comparison to first order perturbation theory (dash-dotted lines).}
  \label{fig:diffconductance2}
\end{figure}
Let us for completeness look at the effect of the gate voltage on the differential conductance. This is shown in \fref{fig:diffconductance2}. The interplay of the gate voltage and bias voltage induced collapse of the main resonance is complicated, but the low-bias behavior is reminiscent of the differential conductance in the SIAM at finite magnetic field \cite{anders08,jakobs10b}: The peak in the differential conductance splits as the gate voltage is increased. The kink at $eV=\omega_0$ appears also at finite gate voltages, but there is an additional step close to $eV=2\omega_0$, which can be related to the merge and subsequent vanishing of the features in the spectral density at $\omega=\pm(\omega_0-eV/2)$.

In first order perturbation theory the kinks at $eV=\omega_0$ are far more pronounced due to the fact that the first term in \eref{eq:diffconductance} vanishes, but the steps at $eV=2\omega_0$ are absent. The latter is then clearly a higher order effect. Overall, FRG and perturbation theory agree over the whole range of bias voltages as long as $U_\epe/\Gamma\ll1$.

\section{Conclusions}\label{sec:conclusions}

We have studied the spectral and transport properties of the Anderson--Holstein model both in and out of equilibrium using the functional renormalization group. We have identified the parameter regimes in which our approach is reliable by studying the renormalization of the quantum dot level width and comparing to analytical estimates. We have found clear signatures of the phonon mode in the transport properties: The width of the linear conductance peak as a function of gate voltage is decreased by increasing the coupling strength to the local phonon mode. The differential conductance at finite bias voltages exhibits a kink when the bias voltage equals the phonon frequency, and, away from the particle--hole symmetric point, also when the bias voltage equals twice the phonon frequency.

The thermoelectric properties of the SIAM with an attractive interaction between the dot electrons have been predicted in an NRG study to differ significantly from those of the model with a repulsive interaction \cite{andergassen11}. Since a sufficiently strong interaction with a local phonon mode gives rise to an overall effective attractive interaction between the electrons, it is a natural extension to our work to next study the thermoelectric properties of the Anderson--Holstein model.

\ack
M A Laakso acknowledges the financial support from The Finnish Cultural Foundation. D M Kennes and V Meden are supported by the DFG via FOR723. We thank S Andergassen and T A Costi for fruitful discussions.

\section*{References}
\providecommand{\newblock}{}

\end{document}